\documentclass[12pt]{article}
\usepackage{graphics}
\usepackage{epsfig}

\begin{document}

\setlength{\textheight}{240mm}
\voffset=-15mm
\baselineskip=20pt plus 2pt
\renewcommand{\arraystretch}{1.6}

\begin{center}

{\large \bf  Thermodynamical Properties and Quasi-localized Energy of 
the Stringy Dyonic Black Hole Solution}\\
\vspace{5mm}
\vspace{5mm}
I-Ching Yang$^{\dag \S} $ \footnote{E-mail:icyang@nttu.edu.tw}
Bai-An Chen$^{\ddag}$, and Chung-Chin Tsai$^{\dag}$

$^{\S}$ Systematic and Theoretical Science Research Group \\
and $^{\dag}$Department of Applied Science, \\
National Taitung University, Taitung 95002, Taiwan (R.O.C.)\\
and \\
$^{\ddag}$Department of Physics, National Kaohsiung Normal University, \\
Kaohsiung 82446, Taiwan (R.O.C.) \\

\end{center}
\vspace{5mm}

\begin{center}
{\bf ABSTRACT}
\end{center}

In this article, we calculate the heat flux passing through the horizon $\left. {\bf TS} 
\right|_{r_h}$ and the difference of energy between the Einstein and M{\o}ller prescription 
within the region ${\cal M}$, in which is the region between outer horizon ${\cal H}_+$ and 
inner horizon ${\cal H}_-$, for the modified GHS solution, KLOPP solution and CLH solution.  
The formula $\left. E_{\rm Einstein} \right|_{\cal M} = \left. E _{\rm M{\o}ller} \right|_{\cal M} 
- \sum_{\partial {\cal M}} {\bf TS}$ is obeyed for the mGHS solution and the KLOPP solution, 
but not for the CLH solution.  Also, we suggest a RN-like stringy dyonic black hole solution, 
which comes from the KLOPP solution under a dual transformation, and its thermodynamical 
properties are the same as the KLOPP solution.

\vspace{2mm}
\noindent
{PACS No.:04.60.Cf, 04.70.Dy.  \\}

\vspace{5mm}
\noindent

\newpage

\section{Introduction}

Since the discovery of black hole evaporation in 1974~\cite{H} , Hawking radiation is 
an important and outstanding quantum effect arising from the quantization of matter 
fields in a background space-time with an event horizon.  In order to study quantum 
effects near the event horizon of black hole, a scientific model which unify quantum 
mechanics with general relativity must be considered.  Until now, superstring theories 
are still the most interesting candidates for a consistent quantum theory of gravity.  
From heterotic string theory, the four-dimensional low-energy effective action~\cite{GHS} 
in which gravity is coupled to electromagnetic and dilaton field is 
obtained as 
\begin{equation}
S = \int d^4 x \sqrt{-g} \left[ - R + 2 (\nabla \phi)^2 + e^{-2\phi} F^2 \right]  .
\end{equation}
With respect to the $U(1)$ potential $A_{\mu}$, $\phi$ and $g_{\mu \nu}$, the field 
equations of the effective action (1) are 
\begin{eqnarray}
&& \nabla_{\mu}( e^{-2\phi} F^{\mu \nu} ) = 0 , \\
&& \nabla^2 \phi + \frac{1}{2} e^{-2\phi} F^2 = 0 , \\
&& R_{\mu \nu} = 2 \nabla_{\mu} \phi \nabla_{\nu} \phi + 2 e^{-2\phi} F_{\mu \rho} 
F_{\nu}^{\rho} - \frac{1}{2} g_{\mu \nu} e^{-2\phi} F^2 .
\end{eqnarray}

In 1991, a magnetically charged black hole solution of field equations found by Garfinkle, 
Horowitz and Strominger (GHS solution)~\cite{GHS} is 
\begin{eqnarray}
&& ds^2 = (1 - \frac{2M}{r}) dt^2 - (1 - \frac{2M}{r})^{-1} dr^2 - r (r+2 \Theta ) d\Omega , \\
&& e^{-2\phi} =  e^{-2\phi_0} \left( 1+ \frac{2 \Theta}{r} \right) , \\
&& F = Q \sin \theta d\theta \wedge d\varphi  ,
\end{eqnarray}
with mass $M$, magnetic charge $Q$ and dilaton charge $\Theta = -Q^2 e^{-2\phi_0} /2M$.
Here $\phi_0$ is the asymptotic constant values of the dilaton.
Later, two distinct stringy dyonic black hole solutions are showed by Kallosh {\it et. al.}~\cite{KLOPP} 
using the line element 
\begin{equation}
ds^2 = C dt^2 - \frac{dr^2}{C} -D^2 r^2 d\Omega , 
\end{equation}
and by Cheng {\it et. al.}~\cite{CLH} using the line element
\begin{equation}
ds^2 = A dt^2 - \frac{B}{A} d \bar{r}^2 - \bar{r}^2 d\Omega  .
\end{equation}
The relation between Eq. (8) and (9) is under a radial coordinate transformation 
\begin{equation}
r = \int \sqrt{B} d \bar{r}  .
\end{equation}
However, the different choices of coordinates of black hole will present different structures
of space-time.  Through the article we use Planck units ($G=  c = k_B = \hbar =1$), and 
follow the convention that Latin indices run from 1 to 3 and Greek indices run from 0 to 3.  
In this article, we will investigate the thermodynamical properties and the quasi-localized 
energy complexes of those stringy dyonic black hole solutions.

\section{Stringy Dyonic Black Hole Solutions}

Subsequently, based on Eq. (8),  Kallosh, Linde, Ort\'{i}n, Peet and van Proeyen presented 
a stringy dyonic black hole solution (KLOPP solution) in the following formulas~\cite{KLOPP}
\begin{eqnarray}
ds^2 & = &  e^{2U} dt^2 - e^{-2U} d{r^*}^2 - R^2 d\Omega, \\
e^{ 2\phi} & = &  e^{2\phi_0} \frac{r^* +\Sigma}{r^* -\Sigma} , \\
F & = & \frac{Q e^{\phi_0}}{(r^* -\Sigma )^2} dt \wedge dr^*  , \\
G & = &  \frac{P e^{-\phi_0}}{(r^* +\Sigma )^2} dt \wedge dr^*  ,
\end{eqnarray}
where
\begin{eqnarray}
e^{2U} & = &  \frac{(r^* - r^*_+ )(r^* - r^*_- )}{R^2} , \\
R^2 & = & {r^*}^2 -\Sigma^2  ,  \\
\Sigma & = & \frac{P^2 - Q^2}{2M} , \\
r^*_{\pm} & = & M \pm \sqrt{M^2 - \Gamma + \Sigma^2}  ,\\
\Gamma & = & P^2 + Q^2 .
\end{eqnarray}
In particular, the same as GHS solution, the area will go to zero as $r^*=\pm \Sigma$ 
causing this surface to be singular.  Here, GHS solution would be a special case 
of KLOPP solution when we introduce a coordinate transformation $ r^* = r +\rho $, 
and set electric charge $Q=0$ and magnetic charge $P=Q_m e^{\phi_0}$ in which 
$\Sigma = -\Theta$. 

Furthermore, another stringy dyonic black hole solution is found by Cheng, Lin
and Hsu (CLH solution) using Eq. (9) in the form~\cite{CLH} 
\begin{eqnarray}
ds^2 & = & \Delta^2 dt^2 - \frac{\sigma^2}{\Delta^2} d\bar{r}^2 - \bar{r}^2 d\Omega, \\
e^{ 2\phi} & = &  e^{2\phi_0} \left( 1- \frac{2\rho}{\sqrt{\bar{r}^2+\rho^2}} +\rho \right) , \\
F_{01} & = & \frac{Q_e}{\bar{r}^2} e^{2\phi_0} , \\
F_{23} & = &  \frac{Q_m}{\bar{r}^2} ,
\end{eqnarray}
where
\begin{eqnarray}
\sigma^2 & = & \frac{\bar{r}^2}{\bar{r}^2 + \rho^2}  , \\
\Delta^2 & = & 1 - \frac{2M}{\bar{r}^2}\sqrt{\bar{r}^2 +\rho^2} + \frac{\beta}{\bar{r}^2} , \\
\rho & = & \frac{1}{2M} \left( Q_e^2 e^{2\phi_0} - Q_m^2 e^{-2\phi_0} \right) , \\
\beta & = & \left( Q_e^2 e^{2\phi_0} + Q_m^2 e^{-2\phi_0} \right)  .
\end{eqnarray}
Dissimilarly, CLH solution will not give zero surface area with non-zero radius, and has two event 
horizons at  $\bar{r}_{\pm} = \left[ (2M^2 -\beta) \pm 2M \sqrt{M^2 -\beta + \rho^2} \right]^{1/2}$.
At the same time, by using a coordinate transformation  $\bar{r}^2 = r^2 + 2\rho r$, CLH solution 
will reduce to GHS solution as setting electric charge $Q_e =0$ and magnetic charge $Q_m=Q$, 
in which $\rho = \Theta$ and $\beta=-2M \Theta $.

By those two stringy dyonic black hole solutions, Eqs. (11)-(14) and Eqs. (20)-(23), the GHS 
solution can be modified to a dyonic black hole solution (mGHS solution) as
\begin{equation}
ds^2 = \Xi dt^2 - \Xi^{-1} dr^2 - \bar{R}^2 d\Omega  ,
\end{equation}
where
\begin{eqnarray}
\Xi & = & \frac{(r- r_+)(r- r_-)}{\bar{R}^2}  , \\
\bar{R}^2 & = & r (r+ 2\rho)  , \\
r_{\pm} & = & M - \rho \pm \sqrt{M^2 -\beta +\rho^2}  .
\end{eqnarray}
The parameters $\beta$ and $\rho$ of above equation are defined in Eqs. (26)-(27).
Note that the modified GHS solution also has two event horizon at $r_+$ and $r_-$.

\section{Thermodynamics of Stringy Black Hole}
In order to investigate the thermodynamical properties of black hole, we need to calculate 
the temperature ${\bf T}$ and entropy ${\bf S}$ of black hole which are given as
\begin{eqnarray}
{\bf T} & = & \frac{\kappa}{2\pi}  , \\
{\bf S} & = & \frac{{\cal A}}{4}  ,
\end{eqnarray}
where the surface gravity $\kappa$~\cite{V} and area ${\cal A}$~\cite{H} is 
\begin{eqnarray}
\kappa & = & \lim_{r \rightarrow r_H} \left( \frac{1}{2} \frac{\partial_r g_{tt}}{\sqrt{g_{tt} g_{rr}}} 
\right)  ,  \\
{\cal A} & = & \int \sqrt{g_{\theta \theta} g_{\phi \phi}}  d\theta d\phi  .
\end{eqnarray}
Also, we study the heat flux passing the event horizon of black hole.  Hence, the 
temperature, entropy and heat flux passing the event horizon of foresaid three black hole 
solution are obtained as : \\
(i) modified GHS solution 
\begin{eqnarray}
{\bf T} & = & \frac{r_h -r_-}{4\pi \bar{R}_h^2} +\frac{r_h  -r_+}{4\pi \bar{R}_h^2} 
-\frac{(r_h +\rho) (r_h  -r_+)(r_h -r_-)}{2\pi \bar{R}_h^4} ,\\
{\bf S} & = & \pi \bar{R}_h^2 , \\
\left. {\bf TS} \right|_{r_h} & = & \frac{r_h - r_+}{4} + \frac{r_h - r_-}{4} 
- \frac{(r_h + \rho)(r_h - r_+)(r_h - r_+)}{2 r_h ( r_h + 2\rho )}  ,
\end{eqnarray}
where $r_h$ is its event horizon.  \\
(ii) KLOPP solution
\begin{eqnarray}
{\bf T} & = & \frac{r^*_h -r^*_-}{4\pi R_h^2} +\frac{r^*_h -r^*_+}{4\pi R_h^2} 
-\frac{r^*_h (r^*_h  -r^*_+)(r^*_h -r^*_-)}{2\pi R_h^4 } ,\\
{\bf S} & = & \pi R_h^2  , \\
\left. {\bf TS} \right|_{r^*_h} & = & \frac{r^*_h - r^*_+}{4} + \frac{r^*_h  - r^*_-}{4} 
- \frac{r^*_h (r^*_h  - r^*_+)(r^*_h  - r^*_+)}{2 \left( {r^*_h}^2 -\Sigma^2 \right)}  ,
\end{eqnarray}
where $r^*_h$ means its event horizon.  \\
(iii) CLH solution
\begin{eqnarray}
{\bf T} & = & \frac{M}{2\pi \bar{r}_h^2} + \frac{M \rho^2}{\pi \bar{r}_h^4} 
-\frac{\beta}{2\pi \bar{r}_h^4} \sqrt{\bar{r}_h^2 +\rho^2} ,\\
{\bf S} & = & \pi \bar{r}_h^2 ,  \\
\left. {\bf TS} \right|_{\bar{r}_h} & = & \frac{M}{2} +\frac{M \rho^2}{\bar{r}_h^2} 
- \frac{\beta}{2 \bar{r}_h^2} \sqrt{\bar{r}_h^2 +\rho^2} ,
\end{eqnarray}
where $\bar{r}_h$ typify its event horizon, and satisfy that 
$2M \sqrt{\bar{r}_h^2 +\rho^2} = \bar{r}_h^2 +\beta$.   So, the temperature and heat flux
can be replaced as 
\begin{eqnarray}
{\bf T} & = & \frac{M}{2\pi \bar{r}_h^2} + \frac{M \rho^2}{\pi \bar{r}_h^4} 
-\frac{\beta}{4\pi M \bar{r}_h^2} -\frac{\beta^2}{4\pi M \bar{r}_h^4} ,\\
\left. {\bf TS} \right|_{\bar{r}_h} & = & \frac{M}{2} +\frac{M \rho^2}{\bar{r}_h^2} 
-\frac{\beta}{4M} -\frac{\beta^2}{4M \bar{r}_h^2}  .
\end{eqnarray}
These results of KLOPP solution are very similar 
to GHS solution, but CLH solution not.

\section{Four Kinds of Energy Complexes}
Next, let us to consider the energy complex in the Einstein~\cite{E}, Weinberg~\cite{W}, 
Landau-Lifshitz~\cite{LL} and M{\o}ller\cite{M} prescription.  The energy component of 
the Einstein energy-momentum complex~\cite{E} is given by
\begin{equation}
E_{\rm Einstein} = \frac{1}{16\pi} \int \frac{\partial H_0^{\;\;0i}}{\partial x^i} d^3x ,
\end{equation}
where 
\begin{equation}
H_0^{\;\;0i} = \frac{g_{00}}{\sqrt{-g}} \frac{\partial}{\partial x^m} \left[ (-g) g^{00} g^{im} \right] .
\end{equation}
Applying Gauss's theorem one obtain
\begin{equation}
E_{\rm Einstein} = \frac{1}{16\pi} \oint H_0^{\;\;0i} \hat{n}_i \cdot d\vec{S} ,
\end{equation}
where $\hat{n}_i$ is the outward unit normal vector over the infinitesimal surface element 
$d\vec{S}$.  In the same way, the energy component of the Weinberg energy-momentum 
complex~\cite{W} within radius $r$ is given by
\begin{equation}
E_{\rm Weinberg} = \frac{1}{16\pi} \oint Q^{i00} \hat{n}_i \cdot d\vec{S} ,
\end{equation}
where
\begin{equation}
Q^{i00} = \frac{\partial h_{jj}}{\partial x^i} - \frac{\partial h_{ij}}{\partial x^j}  ,
\end{equation}
and of the Landau-Lifshitz energy-momentum complex~\cite{LL} 
\begin{equation}
E_{\rm Landau-Lifshitz} =  \frac{1}{16\pi} \oint \frac{\partial \lambda^{00ik}}{\partial x^k} 
\hat{n}_i \cdot d\vec{S}  ,
\end{equation}
where
\begin{equation}
\lambda^{00ik} = (-g) \left( g^{00} g^{ik} \right)  .
\end{equation}
According to the definition of the M{\o}ller energy-momentum complex~\cite{M} and 
Gauss's theorem, the energy component is given as 
\begin{equation}
E_{\rm M{\o}ller} = \frac{1}{8\pi} \oint \chi_0^{\;\;0i} \hat{n}_i \cdot d\vec{S} ,
\end{equation}
where 
\begin{equation}
\chi^{0i}_0 =  \sqrt{-g} \left( - \frac{\partial g_{00}}{\partial x^i} \right) g^{00} g^{ii}  .
\end{equation}

Beginning with the modified GHS solution and the KLOPP solution, their line element 
is like Eq. (8)  as
\begin{equation}
ds^2 =C dt^2 - \frac{dr^2}{C} - D r^2 ( d\theta^2 + \sin^2 \theta d \varphi^2 ).
\end{equation}
The energy component within radius $r$ obtained by the Einstein complex is 
\begin{equation}
E_{\rm Einstein} = \frac{r}{2} (1 - CD) - \frac{r^2 C}{2} \frac{dD}{dr}  ,
\end{equation}
by the Weinberg complex is
\begin{equation}
E_{\rm Weinberg} = \frac{r}{2} (\frac{1}{C} - D) - \frac{r^2}{2} \frac{dD}{dr}  ,
\end{equation}
and by the Landau-Lifshitz energy complex is
\begin{equation}
E_{\rm Landau-Lifshitz} =  \frac{r D}{2} (\frac{1}{C} - D) - \frac{r^2 D}{2} \frac{dD}{dr}  .
\end{equation}
For the modified GHS solution, the relation of energy component of the Einstein, Weinberg 
and Landau-Lifshitz  complex is 
\begin{equation}
E_{\rm Einstein}^{(\rm mGHS)} = \Xi E_{\rm Weinberg}^{(\rm mGHS)} =  
\frac{r^2 \Xi}{\bar{R}^2} E_{\rm Landau-Lifshitz}^{(\rm mGHS)}  ,
\end{equation}
and in the Einstein prescription, the expression of energy within radius $r$  is
\begin{equation}
E_{\rm Einstein}^{(\rm mGHS)} = \frac{r_+ + r_-}{2} - \frac{r_+ r_-}{2r}  
+ \frac{\rho (r- r_+)(r- r_-)}{2 \bar{R}^2} .
\end{equation} 
Similarly, for the KLOPP solution, the relation between these three energy component is 
\begin{equation}
E_{\rm Einstein}^{(\rm KLOPP)} = e^{2U} E_{\rm Weinberg}^{(\rm KLOPP)} =  
\frac{{r^*}^2 e^{2U}}{R^2} E_{\rm Landau-Lifshitz}^{(\rm KLOPP)}  ,
\end{equation}
and the expression of energy in the Einstein prescription is
\begin{equation}
E_{\rm Einstein}^{(\rm KLOPP)} = \frac{r^*_+ + r^*_-}{2} - \frac{r^*_+ r^*_-}{2r^*}  
- \frac{\Sigma^2 (r^* - r^*_+)(r^* - r^*_-)}{r^* R^2} .
\end{equation} 
On the other hand, the energy component with radius $r$ obtained using the M{\o}ller 
complex is
\begin{equation}
E_{\rm M{\o}ller} = \frac{r^2}{2} D \frac{dC}{dr} ,
\end{equation}
and its energy values of the modified GHS solution and the KLOPP solution are
\begin{eqnarray}
E_{\rm M{\o}ller}^{(\rm mGHS)} & = & \frac{r - r_+}{2} + \frac{r - r_-}{2}  
- \frac{(r + \rho)(r - r_+)(r - r_-)}{\bar{R}^2}  , \\
E_{\rm M{\o}ller}^{(\rm KLOPP)} & = & \frac{r^* - r^*_+}{2} + \frac{r^* - r^*_-}{2}  
- \frac{r^* (r^*  - r^*_+)(r^*  - r^*_-)}{ R^2} .
\end{eqnarray}

In the case of the CLH solution , its line element is given as Eq. (9)
\begin{equation}
ds^2 = A dt^2 - \frac{B}{A} d\bar{r}^2 - \bar{r}^2 ( d\theta^2 + \sin^2 \theta d \varphi^2 )  .
\end{equation}
Thus, the energy component within radius $\bar{r}$ obtained by the Einstein complex is  
\begin{equation}
E_{\rm Einstein} = \frac{\bar{r} A}{2 \sqrt{B}} (\frac{B}{A} -1)  ,
\end{equation}
by the Weinberg complex is 
\begin{equation}
E_{\rm Weinberg} = \frac{\bar{r}}{2} (\frac{B}{A} -1)  ,
\end{equation}
by the Landau-Lifshitz complex is 
\begin{equation}
E_{\rm Landau-Lifshitz} = \frac{\bar{r}}{2} (\frac{B}{A} -1)  ,
\end{equation}
and by the M{\o}ller complex is
\begin{equation}
E_{\rm M{\o}ller} = \frac{\bar{r}^2}{2 \sqrt{B}}  \frac{dA}{d\bar{r}}  .
\end{equation}
The relation of energy component of the Einstein, Weinberg and Landau-Lifshitz 
complex is 
\begin{equation}
E_{\rm Einstein}^{(\rm CLH)} = \frac{\Delta^2}{\sigma} E_{\rm Weinberg}^{(\rm CLH)} =  
\frac{\Delta^2}{\sigma} E_{\rm Landau-Lifshitz}^{(\rm CLH)}  .
\end{equation}
In previous studies of Yang {\it et. al.}~\cite{Y9X}, the expression of Einstein's energy complex within 
radius $\bar{r}$ is 
\begin{equation}
E_{\rm Einstein}^{(\rm CLH)} = M + \frac{M \rho^2}{\bar{r}^2} 
- \frac{\beta \sqrt{\bar{r}^2 +\rho^2}}{2\bar{r}^2} - \frac{\rho^2}{2 \sqrt{\bar{r}^2 +\rho^2}}   ,
\end{equation}
and of M{\o}ller's energy complex is 
\begin{equation}
E_{\rm M{\o}ller}^{(\rm CLH)} = M + \frac{2M \rho^2}{\bar{r}^2} 
- \frac{\beta \sqrt{\bar{r}^2 +\rho^2}}{\bar{r}^2} .
\end{equation}

\section{Quasi-localized Energy Complexes and Thermodynamical Potentials}
Moreover, the difference of energy between the Einstein and M{\o}ller prescription~\cite{Y04} 
is defined as 
\begin{equation}
\Delta E = E_{\rm Einstein} - E_{\rm M{\o}ller}  ,
\end{equation}
and its values for the mGHS, KLOPP and CLH solution are 
\begin{eqnarray}
\Delta E^{\rm (mGHS)} & = & (r_+ + r_-) - \left( r + \frac{r_+ r_-}{2r} \right)  \\
 & & + \frac{(r - r_+)(r - r_-)}{\bar{R}^2} \left( r + 2\rho \right)  , \\
\Delta E^{\rm (KLOPP)} & = & (r^*_+ + r^*_-) - \left( r^* + \frac{r^*_+ r^*_-}{2r^*} \right)  \\
 & & + \frac{(r^* - r^*_+)(r^* - r^*_-)}{R^2} \left( r^* - \frac{\Sigma^2}{r^*} \right)  , \\
\Delta E^{\rm (CLH)} & = &   \frac{\beta \sqrt{\bar{r}^2 + \rho^2}}{2 \bar{r}^2} - \frac{M \rho^2}{\bar{r}^2}
- \frac{\rho^2}{2 \sqrt{\bar{r}^2 + \rho^2}}  .
\end{eqnarray}
In the mGHS, KLOPP and CLH solutions, they are the existence of two different  Cauchy horizons, 
event horizon ${\cal H}_+$ is located at $r=r_+$ and inner horizon ${\cal H}_-$ is located at $r=r_-$.   
So, we can suppose that ${\cal M}= \left\{ (t, r, \theta, \phi ) | r_+ > r > r_- \right\} $ is the region 
between ${\cal H}_+$ and ${\cal H}_-$.  However, one suggest that the temperature of the inner Cauchy 
horizon must be defined as~\cite{Z}
\begin{equation}
\left. {\bf T} \right|_{r_-} \equiv - \left. \frac{\kappa}{2\pi} \right|_{r_-}  .
\end{equation}
Because of 
\begin{eqnarray}
& & \left. \Delta E^{\rm (mGHS)} \right|^{r_+}_{r_-} = - \left( \frac{r_+ - r_-}{2} \right)  ,   \\
& & {\bf TS} |_{r_+} = \frac{r_+ - r_-}{4} = {\bf TS} |_{r_-} ,
\end{eqnarray}
for the mGHS solution, we have 
\begin{equation}
\left. E^{\rm (mGHS)}_{\rm Einstein} \right|_{\cal M} = \left. E^{\rm (mGHS)} _{\rm M{\o}ller} \right|_{\cal M} 
- \sum_{\partial {\cal M}} {\bf TS}  .
\end{equation}
Similarly, base on these results
\begin{eqnarray}
& & \left. \Delta E^{\rm (KLOPP)} \right|^{r^*_+}_{r^*_-} = - \left( \frac{r^*_+ - r^*_-}{2} \right) ,  \\
& & {\bf TS} |_{r^*_+} = \frac{r^*_+ - r^*_-}{4} = {\bf TS} |_{r^*_-} ,
\end{eqnarray}
we also can obtain
\begin{equation}
\left. E^{\rm (KLOPP)}_{\rm Einstein} \right|_{\cal M} = \left. E^{\rm (KLOPP)} _{\rm M{\o}ller} \right|_{\cal M} 
- \sum_{\partial {\cal M}} {\bf TS}  .
\end{equation}
Base on the viewpoint of Nester {\it et. al.}~\cite{CNC}, we could suggest $E_{\rm Einstein}$ 
and $E_{\rm M{\o}ller}$ as thermodynamic potentials, because Eq. (84) and Eq. (87) are like 
the Legendre transformation.  Furthermore, for the CLH solution, the heat fluxes on those two 
Cauchy horizons are 
\begin{eqnarray}
{\bf TS} |_{\bar{r}_+} = \frac{\bar{r}_+^2}{4M}  ,  \\
{\bf TS} |_{\bar{r}_-} = - \frac{\bar{r}_-^2}{4M}  ,
\end{eqnarray}
and the difference of energy in the region ${\cal M}$ is 
\begin{equation}
\left. \Delta E^{\rm (CLH)} \right|^{\bar{r}_+}_{\bar{r}_-} = 
 \frac{\beta (\bar{r}_-^2 - \bar{r}_+^2)}{4M \beta +4M \rho^2}  ,
\end{equation}
Here, we will obtain that 
\begin{equation}
\left. E^{\rm (CLH)}_{\rm Einstein} \right|_{\cal M} = \left. E^{\rm (CLH)} _{\rm M{\o}ller} \right|_{\cal M} 
 - \frac{\beta}{\beta +\rho^2} \sum_{\partial {\cal M}} {\bf TS}  ,
\end{equation}
and can not get the same result, like Eq. (83) and (86), except $\rho =0$. 

\section{Conclusions}

In this article, we have calculated two kinds of physical properties, thermodynamical and 
mechanical properties, and have obtained the results of both properties about those three 
stringy black hole solutions.  The thermodynamical properties include the temperature ${\bf T}$, 
the entropy ${\bf S}$ and the heat flux passing through the horizon $\left. {\bf TS} \right|_{r_h}$,
and the mechanical properties include four kinds of energy complexes $E_{\rm Einstein}$, 
$E_{\rm Weinberg}$, $E_{\rm Landau-Lifshitz}$ and $E_{\rm M{\o}ller}$, and the difference 
of energy between the Einstein and M{\o}ller prescription $\Delta E$.  For mGHS solution 
and KLOPP solution,  particularly, the difference of energy between the Einstein and M{\o}ller 
prescription within the region ${\cal M}$ is equal to the heat flux which is exhibited on every 
boundary of ${\cal M}$, like as the formula that we have pointed out~\cite{Y09}
\begin{equation}
\left. E_{\rm Einstein} \right|_{\cal M} = \left. E _{\rm M{\o}ller} \right|_{\cal M} 
- \sum_{\partial {\cal M}} {\bf TS}  ,
\end{equation}
and the heat flux passing by the outer horizon ${\cal H}_+$ and by inner horizon ${\cal H}_-$
are the same, like as
\begin{equation}
{\bf TS} |_{{\cal H}_+} = {\bf TS} |_{{\cal H}_-}  .
\end{equation}
However, on the case of CLH solution, because of 
\begin{equation}
{\bf TS} |_{\bar{r}_+}   \neq {\bf TS} |_{\bar{r}_-}  ,
\end{equation}
we would not obtain the same identical equation (92).  The relation of coordinate 
transformation between the mGHS solution and KLOPP solution is a linear transformation 
$r^* = r+ \rho$, but between the KLOPP solution and CLH solution is non-linear 
transformation ${r^*}^2 = \bar{r}^2 +\rho^2$, in which the relation between $r^*$ 
and $\bar{r}$ is like the hyperbolic functions $\cosh^2 x - \sinh^2 x =1$.  Thus, the 
space-time structure of KLOPP solution is homeomorphism with mGHS solution, but
not with CLH solution.  According to this condition (93), there occurs equal heat 
fluxes on both Cauchy horizons of mGHS solution and KLOPP solution.  Then, mGHS 
solution or KLOPP solution is more suitable to study the thermodynamics of stringy 
black hole solution than CLH solution. 
  
On the other hand, the effective action Eq. (1) and field equation Eq. (3) and Eq. (4) 
will be invariant under~\cite{GHS}
\begin{equation}
F_{\mu \nu} \longrightarrow \tilde{F}_{\mu \nu} = 
\frac{1}{2} e^{-2\phi} \epsilon_{\mu \nu}^{\lambda \eta} F_{\lambda \eta} .
\end{equation}
Let us introduce the notation
\begin{eqnarray}
& & \bar{Q} = \frac{1}{\sqrt{2}} \left( Q+P \right)  , \\
& & \bar{P} = \frac{1}{\sqrt{2}} \left( Q-P \right)  ,
\end{eqnarray}
so that
\begin{eqnarray}
& & \Sigma = - \frac{\bar{Q} \bar{P}}{M}  , \\
& & \beta = \Gamma =\bar{Q}^2 + \bar{P}^2 .
\end{eqnarray}
The Eq. (15) in the KLOPP solution will be modified as 
\begin{equation}
e^{2U} = \left( 1 - \frac{2M}{r^*} + \frac{\bar{Q}^2 + \bar{P}^2 - \Sigma^2}{{r^*}^2} \right) 
\frac{{r^*}^2}{R^2},
\end{equation}
This new solution is RN-like black hole solution and its thermal properties is the same as 
KLOPP solution.

\end{document}